 \documentclass[twoside,reqno,11pt]{gocp}

\usepackage{graphicx}
\usepackage{epsfig}
\usepackage{amsthm}
\usepackage{amsmath}
\usepackage{latexsym}
\usepackage{amsfonts}
\usepackage{amssymb}

 \usepackage{hyperref}  

 \def\theequation{\arabic{section}.\arabic{equation}}

 \title[The fractional Schr\"odinger equation  \dots]{
The fractional Schr\"odinger equation \\ [2pt]
and the infinite potential well -\\ [2pt]
 numerical results using the Riesz derivative}
 \author[R. Herrmann]{Richard Herrmann $^1$}

\begin{document}
\noindent
\includegraphics[scale=0.15]{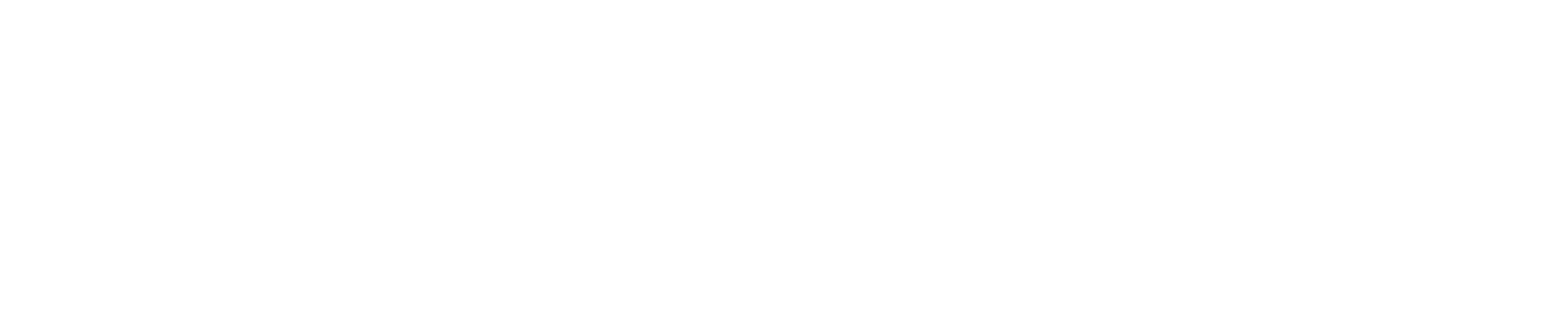}

\bigskip
\bigskip 

 \begin{abstract}
Based on the Riesz definition of the fractional derivative  the fractional Schr\"odinger equation with an infinite well potential is investigated. First it is shown analytically, that the solutions of the free fractional Schr\"odinger equation are not eigenfunctions, but good approximations for large k and for $\alpha \approx 2$. The first lowest eigenfunctions are then calculated numerically and an approximate analytic  formula for the level spectrum is derived.   
 \medskip

 \smallskip

 \end{abstract}

 \maketitle


\section{Introduction}\label{sec:1}
Wave equations play a significant role in the description of the dynamic development of particles and fields; e.g. the Maxwell-equations describe the behavior of the electro-magnetic field in terms of coupled partial differential equations. In quantum mechanics  a particle may be described 
by the non-relativistic   Schr\"odinger wave equation, where the kinetic term is given by the Laplace-operator. 

Fractional calculus introduces the concept of non-locality to arbitrary hitherto local operators \cite{old76}, \cite{pod99}, \cite{her11}. This is a new property, which  only  recently attracted attention on a broader basis. The interest in a non-local dynamic description of e.g.  quantum systems has been steadily increasing, because it is expected, that  quantum phenomena may be treated more elegantly  from a generalized point of view.

Within this context, it is helpful  to investigate fundamental properties of a fractional wave equation and 
to study general  features of its solution.     

As an example, we will present in the following the main results of a solution of the fractional Schr\"o\-di\-nger equation with infinite potential well. 

\setcounter{section}{1}
\setcounter{equation}{0}
\section{The problem - analytic part}
Let the one dimensional fractional stationary Schr\"odinger equation in scaled canonical form be defined as 
\begin{equation}
-\Delta^{\alpha/2} \Psi(x) = (E - V(x)) \Psi(x)
\end{equation}
with the fractional Laplace-operator $\Delta^{\alpha/2}$. 
The definition of a fractional order derivative is not unique,  several definitions 
e.g. the Riemann\cite{rie47}, Caputo\cite{cap67}, Liouville \cite{f2},  Riesz  \cite{riesz}, Feller\cite{feller},  Weyl\cite{weyl}  coexist and are equally well suited for an extension of the standard derivative.

In order to preserve  Hermiticity for the fractional extension of the Laplace-operator\cite{laskin}, we will explicitly apply the Riesz fractional derivative
\begin{eqnarray}
\label{q12driesz}
\Delta^{\alpha/2} \! \equiv \,  _\textrm{\tiny{RZ}}^\infty D^\alpha f(x)\! \! &=&\! \!  \Gamma(1+\alpha)
{\sin(\pi \alpha/2)\over \pi} 
 \int_0^\infty \! 
{f(x+\xi)-2 f(x) + f(x-\xi) \over \xi^{\alpha+1}}d\xi \\
& &  \qquad  \qquad \qquad  \qquad \qquad \qquad  \qquad \qquad  \qquad  \qquad  0< \alpha <2 \nonumber
\end{eqnarray}
where the left superscript in $_\textrm{\tiny{RZ}}^\infty D^\alpha$ emphasizes the fact, that the integral domain is the full space $\mathbb{R}$.

Since the eigenfunctions of the Riesz derivative operator are given as
\begin{eqnarray}
\label{q12dfree1}
_\textrm{\tiny{RZ}}^\infty D^\alpha \cos (k x) &=& -|k|^\alpha  \cos (k x) \\
\label{q12dfree2}
_\textrm{\tiny{RZ}}^\infty D^\alpha \sin (k x) &=& -|k|^\alpha  \sin (k x) 
\end{eqnarray}
the eigenfunctions of the potential free ($V(x)=0)$  fractional  Schr\"odinger equation using the Riesz fractional derivative follow as
\begin{eqnarray}
\label{q12sol}
_\textrm{\tiny{RZ}}\Psi^+(k,x) &=& \cos (k x)  \\
_\textrm{\tiny{RZ}}\Psi^-(k,x) &=& \sin (k x) 
\end{eqnarray}
where the $\pm$-sign indicates the parity and the corresponding continuous energy spectrum follows as:
\begin{equation}
\label{q13efree}
E_k^{\textrm{free}} = |k|^\alpha \quad\quad\quad\quad\quad\quad k \in \mathbb{R}
\end{equation}
Since the integrals  $\int_{-\infty}^{+\infty}dx \cos(kx )$ and  $\int_{-\infty}^{+\infty}dx \sin(kx )$ respectively  are divergent,  
the eigenfunctions are not normalizable on the full domain $\mathbb{R}$. 

In the case $\alpha=2$, which corresponds to the classical quantum mechanics, of course we may apply the classical box-normalization, which means we make a statement on the behavior of the wave-function isolated in a box of size e.g. $2 q$ of the form:
\begin{eqnarray}
\int_{-q}^{+q}  \cos (k x) dx < \infty\\
\int_{-q}^{+q}  \sin (k x) dx < \infty 
\end{eqnarray}
According to (\ref{q12dfree1}) and (\ref{q12dfree2})  a special feature of the Riesz derivative is that  the type of  the eigenfunctions of the free fractional Schr\"odinger equation does not change for arbitrary $\alpha$. Hence the question arises, whether  box-normalization is a legal procedure in the case of non-local differential equations. At least, the physical meaning of box-normalization may be different. 

To avoid these interpretative difficulties we will switch on a potential $V(x)$, such that the eigenfunctions vanish at infinity. An ideal candidate for such a potential is the infinite potential well, centered at the origin with finite size $2 q$, which is explicitly given by: 
\begin{equation}
V(x) = 
\begin{cases}
0         &  \text{$|x| \leq q$}\cr
   \infty &   \text{$|x|>q$}
\end{cases}
\end{equation} 
A reasonable ansatz for the corresponding eigenfunction follows as:
\begin{equation}
\label{fconfined}
^\sqcup\Psi(x) = 
\begin{cases}
^\infty\Psi(x)         &\text{$|x| \leq q$}\cr
   0&   \text{$|x|>q$}
\end{cases}
\end{equation} 
Where the superscript $^\sqcup\Psi(x)$ emphasizes the fact, that the wave-function per definitionem is now  confined inside the infinite potential well and vanishes outside. The superscript $^\infty\Psi(x)$ indicates, that this function may at first be defined on the whole domain, but is used only inside the bounded domain of the infinite potential well. 

The normalization condition for $^\sqcup\Psi(x)$ is now given by:
\begin{equation}
\int_{-\infty}^{+\infty} dx ^\sqcup\Psi(x)     = 
\int_{-q}^{+q} dx^\infty\Psi(x)     =  1
\end{equation} 
As a consequence, we may interpret $^\sqcup\Psi(x)$  physically  as a normalizable  wave-function and its absolute value   $^\sqcup\Psi(x)^\sqcup\Psi(x)^*$ as a probability measure.

In the classical, local case ($\alpha=2$) we obtain immediately
\begin{equation}
^\sqcup\Psi^{\pm}(x) = 
\begin{cases}
\cos{k \frac{\pi}{2 q} x}   &\text{$|x| \leq q$ and  $k = 1,3,5,...$}\cr
\sin{k \frac{\pi}{2 q} x}   &\text{$|x| \leq q$ and  $k = 2,4,6,...$}\cr
   0&   \text{$|x|>q$}
\end{cases}
\end{equation} 
and the continuous energy spectrum changes to a discrete one, since $k \in \mathbb{N}$ is an integer now.
\begin{equation}
E_k^{\textrm{local}}  =  (\frac{\pi}{2 q})^2 k^2  \quad \quad k=1,2,3,...
\end{equation} 

In order to solve the fractional Schr\"odinger equation of the infinite potential well for arbitrary $\alpha$, we have to apply the Riesz derivative operator to ${^\sqcup\Psi(x)}$. 

For the positive semi-axis $x \geq 0$ we obtain for $x \leq q$: 
\begin{eqnarray}
\label{q12d2riesz}
_\textrm{\tiny{RZ}}^\infty D^\alpha ({^\sqcup\Psi(x)}) &=& \Gamma(1+\alpha)
{\sin(\pi \alpha/2)\over \pi} \nonumber \times \\ 
&&   \int_0^\infty
{^\sqcup\Psi(x-\xi) - 2\, {^\sqcup}\Psi(x) + {^\sqcup}\Psi(x+\xi) \over \xi^{\alpha+1}}d\xi \nonumber\\
&=& \Gamma(1+\alpha)
{\sin(\pi \alpha/2)\over \pi} \times \Big\{ \nonumber \\
&&  \int_{0}^{q-x}
{^\infty\Psi(x-\xi)-2 \,^\infty\Psi(x) + {{^\infty\Psi}}(x+\xi) \over \xi^{\alpha+1}}d\xi  + \\
&& \int_{q-x}^{q+x}
{{^\infty\Psi}(x-\xi) -2\, {^\infty\Psi}(x)   \over \xi^{\alpha+1}}d\xi  + \nonumber\\
&& \int_{q+x}^{\infty}
{ -2 \,^\infty\Psi(x)  \over \xi^{\alpha+1}}d\xi  \,\, \Big\} \nonumber\\
&=&
_\textrm{\tiny{RZ}}^\sqcup D^\alpha ({^\infty\Psi(x)}) \quad\quad\quad \quad\quad\quad\quad\quad\quad\quad\quad 0 \leq x\leq q \nonumber
\end{eqnarray}
In addition, for $x > q$ due to the non-local character of the Riesz derivative operator we obtain a non-vanishing finite term
\begin{equation} 
_\textrm{\tiny{RZ}}^\infty D^\alpha ({^\sqcup\Psi(x)}) =\Gamma(1+\alpha)
{\sin(\pi \alpha/2)\over \pi}  \times
 \int_{x-q}^{x+q}{ {^\infty\Psi}(x-\xi) \over \xi^{\alpha+1}}d\xi \quad\quad\quad  x > q 
\end{equation} 
which is negligible only in the special case of the infinite potential well discussed here.

The corresponding  equations  for $x \leq 0$  just interchange the roles of ${^\infty\Psi}(x-\xi) $ and ${^\infty\Psi}(x+\xi) $. Since parity is conserved for the infinite potential well, we may restrict  to (\ref{q12d2riesz}) without loss of generality.

We can write (\ref{q12d2riesz}) in short-hand notation:
\begin{equation}
{_\textrm{\tiny{RZ}}^\infty} D^\alpha ({^\sqcup\Psi(x)}) = 
{_\textrm{\tiny{RZ}}^\sqcup} D^\alpha ({^\infty\Psi(x)}) 
\end{equation}
This may be interpreted as a modification of the Riesz-operator, which now only covers the inside region of the potential well. Obviously  both operators differ significantly.
\begin{equation}
{_\textrm{\tiny{RZ}}^\infty} D^\alpha \neq 
{_\textrm{\tiny{RZ}}^\sqcup} D^\alpha  
\end{equation} 
This is a general feature of all fractional derivative definitions, which span over the full $\mathbb{R}$ e.g. Liouville's, Weyl's and Feller's  definition.

Only for the Riemann ${_\textrm{\tiny{R}}} D^\alpha$  and the Caputo ${_\textrm{\tiny{C}}} D^\alpha$ definition of a fractional derivative and only for the very special case of an infinite  potential well centered at the origin the operator is not altered:
\begin{eqnarray}
{_\textrm{\tiny{R}}^\infty} D^\alpha &=&  
{_\textrm{\tiny{R}}^\sqcup} D^\alpha  \\
{_\textrm{\tiny{C}}^\infty} D^\alpha &=&  
{_\textrm{\tiny{C}}^\sqcup} D^\alpha  
\end{eqnarray} 
because the corresponding derivative definitions  only cover the inside region of the potential well $0\leq x\leq q$.

As a direct consequence, eigenfunctions of the free fractional Schr\"odinger equation based on the Riemann and Caputo derivative definition  automatically determine the eigenfunctions of the same Schr\"odinger equation with infinite potential well fulfilling the additional constraint, that the functions should vanish at the boundaries of the infinite well \cite{her05}.

These eigenfunctions  ${_\textrm{\tiny{R,C}}^\infty}\Psi^{\pm}(x,\alpha)$ are known analytically, normalizable on $\mathbb{R}$ for $\alpha < 2$ and are given in terms of the Mittag-Leffler functions $E_\alpha(z)$ and $E_{\alpha, \beta} (z)$ as:
\begin{eqnarray} 
\label{pot0}
{_\textrm{\tiny{R}}^\infty}\Psi^{+}(x,\alpha) &=&  x^{\frac{\alpha}{2}-1}E_{\alpha,\frac{\alpha}{2}}( -x^{\alpha})\\
{_\textrm{\tiny{R}}^\infty}\Psi^{-}(x,\alpha) &=&  x^{\alpha-1}E_{\alpha, \alpha}( -x^{\alpha})\\
{_\textrm{\tiny{C}}^\infty}\Psi^{+}(x,\alpha) &=&  E_{ \alpha}( -x^{\alpha})\\
{_\textrm{\tiny{C}}^\infty}\Psi^{-}(x,\alpha) &=&  x^\frac{\alpha}{2} E_{\alpha,1+\frac{\alpha}{2}}( -x^{\alpha}) 
\quad\quad\quad\quad 0 \leq \alpha \leq 2
\end{eqnarray} 
which reduce to the trigonometric functions for $\alpha=2$  and the corresponding eigenvalues are determined   from the zeros of these functions  \cite{her05}, \cite{her11}. 
\\

In contrast to this simple classical behavior we now consider the case of the fractional Schr\"odinger equation for the infinite potential well based on  the Riesz definition of a fractional derivative. The following questions arise: 
\begin{itemize}
\item{are plain waves still a solution for the infinite potential well?}
\item{if not, are they at least a good approximation?}
\item{what do the exact solutions look like?}
\end{itemize}
To answer these questions, we rearrange terms in the integral-operators $I_n$ of  the modified Riesz-operator ${_\textrm{\tiny{RZ}}^\sqcup D^\alpha}$:
\begin{equation}
\Big({_\textrm{\tiny{RZ}}^\sqcup D^\alpha} ({^\infty\Psi})\Big)(x)  =
\Big(\Gamma(1+\alpha){\sin(\pi \alpha/2)\over \pi} (I_1 + I_2 +I_3) {^\infty\Psi}\Big)(x)
\end{equation}
where $I_n$ are given explicitely:
\begin{eqnarray}
\label{q12driesz3}
I_1 {\Psi(x)} &=& 
  \int_0^{q-x}
{\Psi(x+\xi) - 2 \Psi(x) + \Psi(x-\xi) \over \xi^{\alpha+1}}d\xi\\
I_2 {\Psi(x)} &=& 
 \int_{q-x}^{q+x}
{ \Psi(x-\xi) \over \xi^{\alpha+1}}d\xi\\
I_3 {\Psi(x)} &=& 
 \int_{q-x}^{\infty}
{ - 2 \Psi(x)  \over \xi^{\alpha+1}}d\xi  =  - 2 \Psi(x) \int_{q-x}^{\infty}
{ 1  \over \xi^{\alpha+1}}d\xi
\end{eqnarray}

\begin{figure}[t]
\begin{center}
\includegraphics[width=130mm]{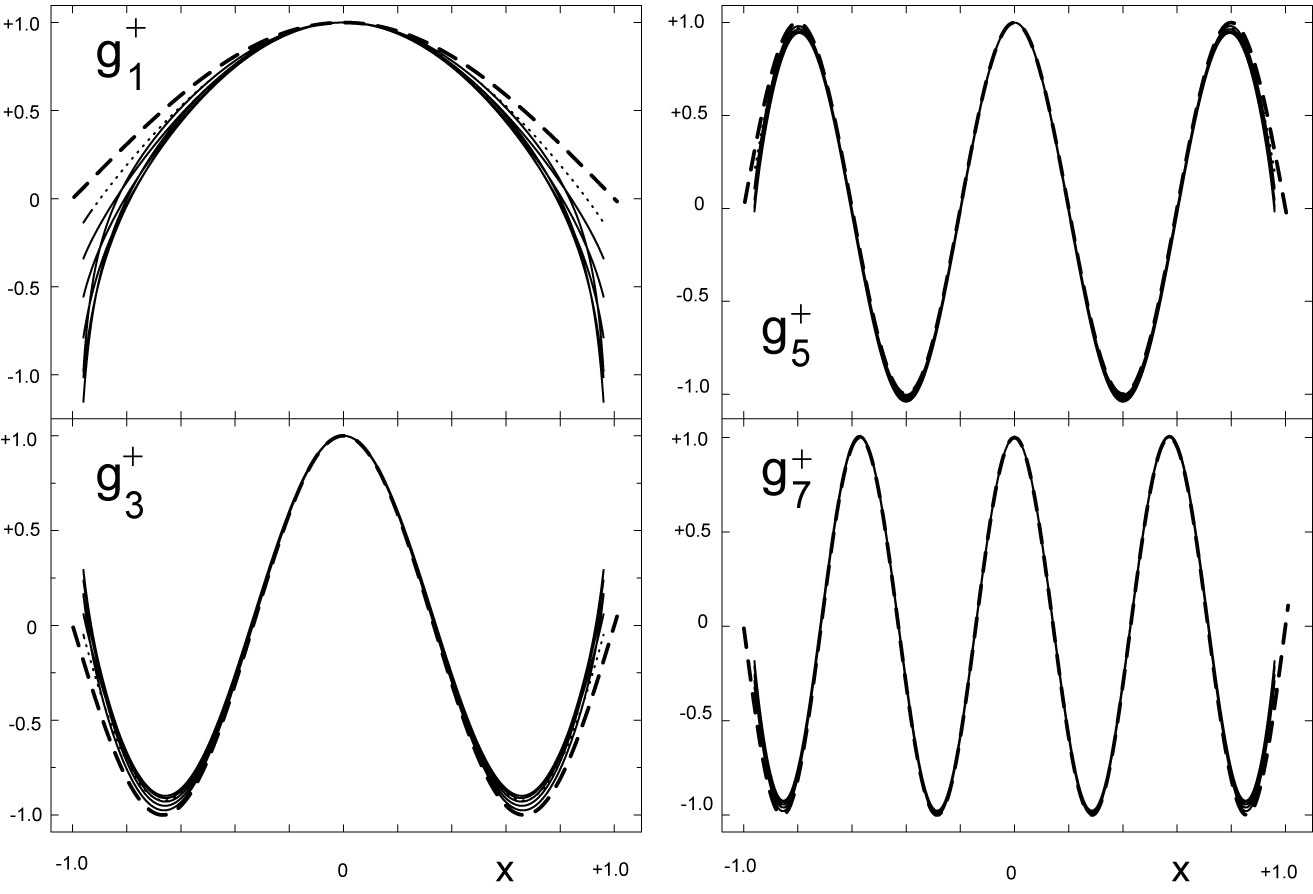}\\
\caption{\label{fcos}
The lowest solutions of $-{_\textrm{\tiny{RZ}}^\sqcup} D^\alpha \cos(k \pi/2 x)= E g_k(\alpha, x)$ for $q=1$ in the range  (dashed line) $2\geq \alpha \geq 0.25$ (dotted line) in $\Delta \alpha = 0.25 $ steps. For the classical, local case $\alpha = 2$ the solution is indeed an eigenfunction, but for decreasing $\alpha$ deviations from $\cos(k \pi/2 x)$ become more and more pronounced.  On the other hand, for large $k$ the error becomes smaller and $\cos(k \pi/2 x)$ becomes a good first guess for the exact eigenfunction. 
} 
\end{center}
\end{figure}

\begin{figure}[t]
\begin{center}
\includegraphics[width=130mm]{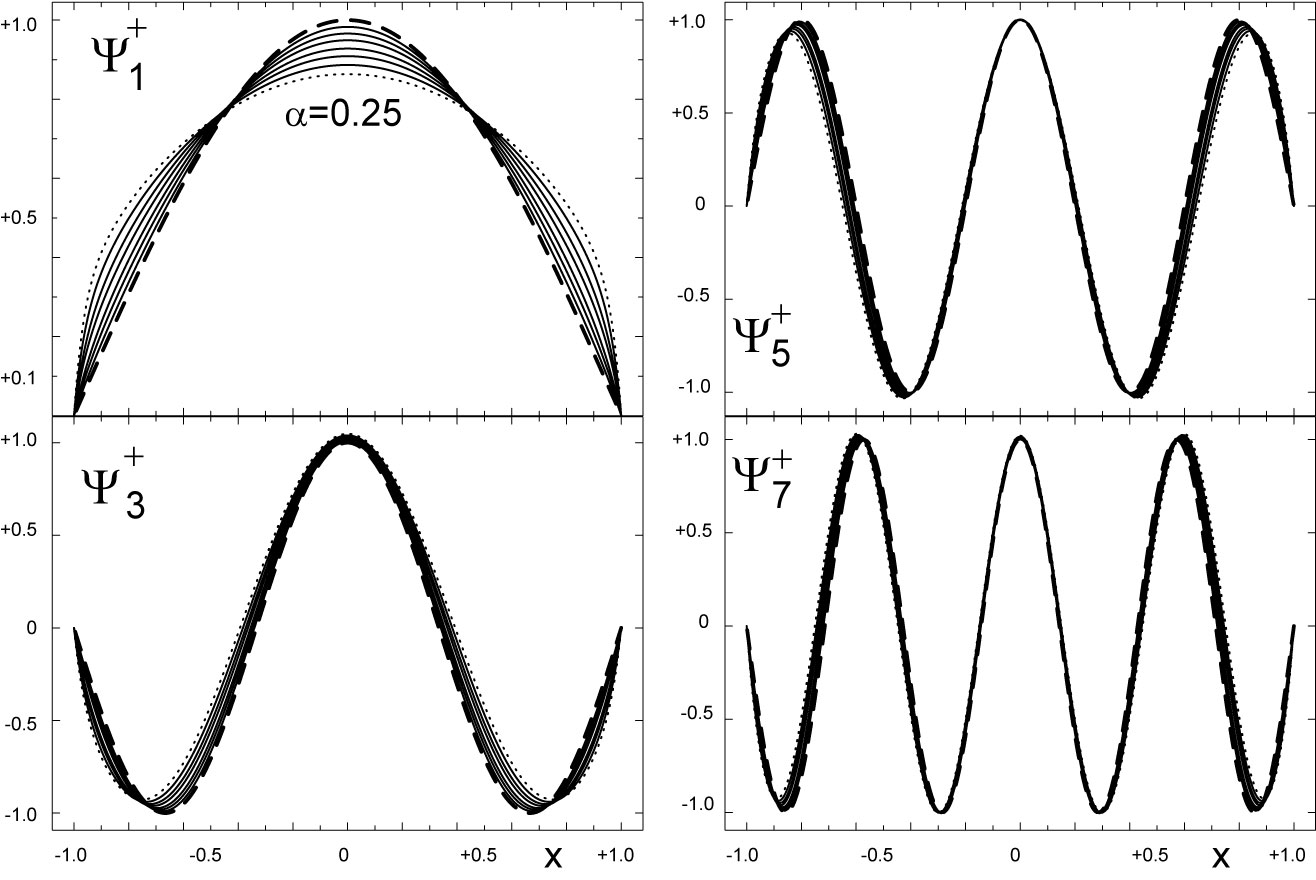}\\
\caption{\label{fcosexact}
The lowest numerically determined  eigenfunctions $\Psi_k^+(\alpha, x)$ for $q=1$ with positive parity  for (dashed line) $2 \geq \alpha \geq 0.25$ (dotted line) in $\Delta \alpha = 0.25 $ steps. 
Only in the limit $\Psi_k^+(\alpha \rightarrow 2, x)= \cos(k \pi/2 x) $ holds. 
} 
\end{center}
\end{figure}

\begin{figure}[t]
\begin{center}
\includegraphics[width=130mm]{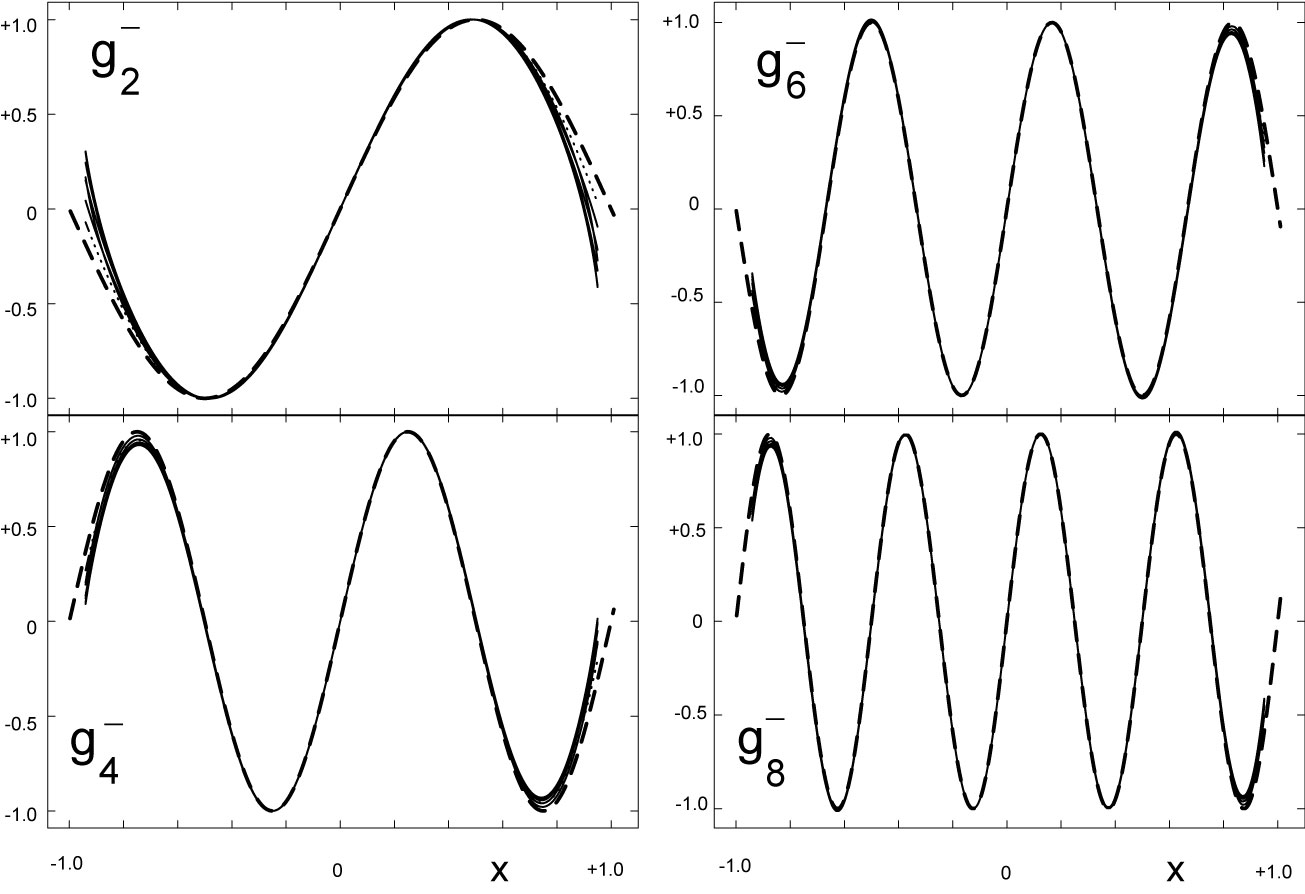}\\
\caption{\label{fsin}
The lowest solutions of $-{_\textrm{\tiny{RZ}}^\sqcup} D^\alpha \sin(k \pi/2 x)= E g_k(\alpha, x)$ for $q=1$ in the range  (dashed line) $2  \geq \alpha \geq 0.25$ (dotted line)  in $\Delta \alpha = 0.25 $ steps. For the classical, local case $\alpha = 2$ the solution is indeed an eigenfunction, but for decreasing $\alpha$ deviations from $\sin(k \pi/2 x)$ become more and more pronounced.  On the other hand, for large $k$ the error becomes smaller and $\sin(k \pi/2 x)$ becomes a good first guess for the exact eigenfunction. 
} 
\end{center}
\end{figure}
\begin{figure}[t]
\begin{center}
\includegraphics[width=130mm]{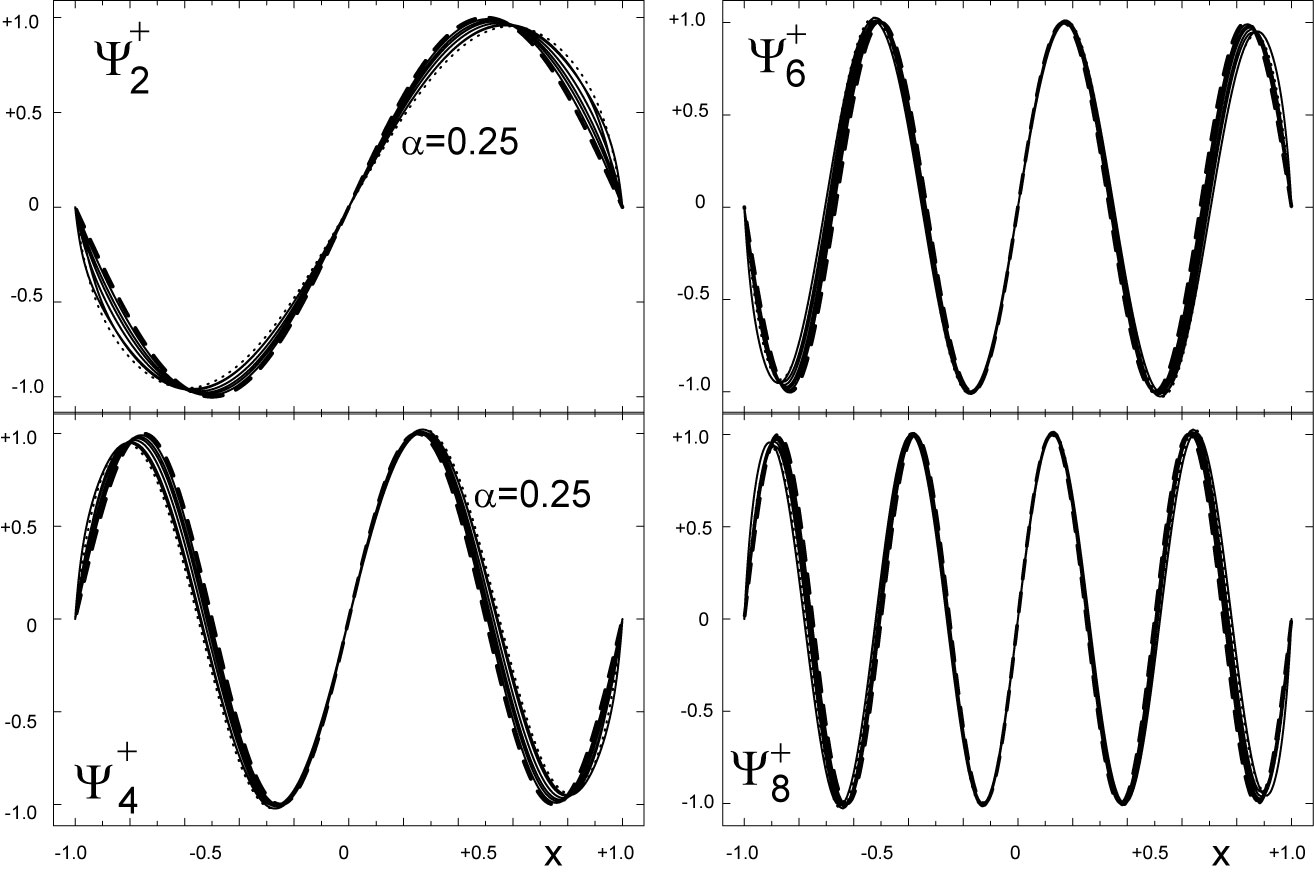}\\
\caption{\label{fsinexact}
The lowest  numerically determined  eigenfunctions $\Psi_k^-(\alpha, x)$ with negative parity for $q=1$ in the range (dashed line) $2 \geq \alpha \geq 0.25$  (dotted line) in $\Delta \alpha = 0.25 $ steps. 
Only in the limit $\Psi_k^+(\alpha \rightarrow 2, x)= \sin(k \pi/2 x) $ holds. 
} 
\end{center}
\end{figure}
\begin{figure}[t]
\begin{center}
\includegraphics[width=130mm]{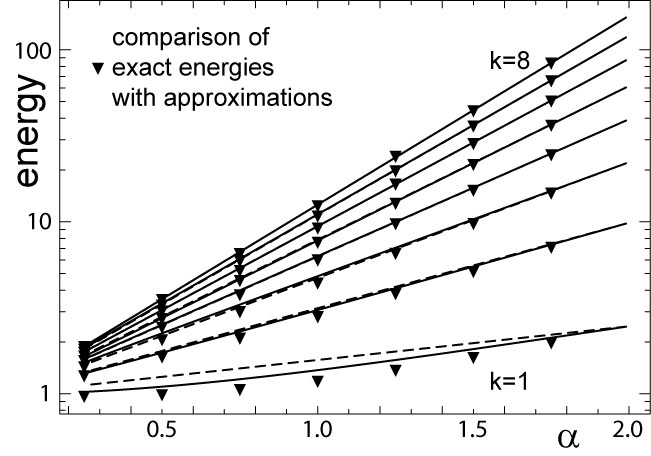}\\
\caption{\label{fenergiesexact}
A comparison of the numerically determined energy  values (triangles) with approximate formulas $E^\sim_k(\alpha)$ from (\ref{etilde}) (solid line) and $E_k^{\textrm{free}}(\alpha)$ from (\ref{q13efree}) (dashed line) normalized to $q=1$.
} 
\end{center}
\end{figure}
Using the test functions $\Psi^+(x) = \cos(k \pi/2 x)$ and $\Psi^-(x) = \sin(k \pi/2 x)$  we obtain with  the help of \cite{Ab}, \cite{erd53} and a cup of tea:
\begin{eqnarray}
\label{q12n3}
I_1 
\begin{cases}
\cos(k \pi/2 x)       & \cr
\sin(k \pi/2 x)       &
\end{cases}
&=& 
{(q-x)^{-\alpha}  \over 2 \alpha }  \Big( \nonumber\\
& &  \frac{k^2 \pi^2}{ \alpha-2} (\frac{q-x}{q})^2  \, _1F_2\big(1-\frac{\alpha}{2};\frac{3}{2}, 2-\frac{\alpha}{2}; -k^2 \pi^2 (\frac{q-x}{4q})^2\big) +\nonumber\\
& & 8 \sin^2( k \pi \frac{q-x}{4q}) \Big) 
\times
\begin{cases}
\cos(k \pi/2 x)       & \cr
\sin(k \pi/2 x)       &
\end{cases}
\\
I_2
\begin{cases}
\cos(k \pi/2 x)       & \cr
\sin(k \pi/2 x)       &
\end{cases}
 &=& 
(\frac{k \pi}{2 q})^{\alpha}
\times
\begin{cases}
\textrm{Re} (\Upsilon(x))       & \cr
-\textrm{Im} (\Upsilon(x))     &
\end{cases}
\\
I_3
\begin{cases}
\cos(k \pi/2 x)       & \cr
\sin(k \pi/2 x)       &
\end{cases}
&=& 
{ -2 (q-x)^{-\alpha} \over  \alpha}
\times
\begin{cases}
\cos(k \pi/2 x)       & \cr
\sin(k \pi/2 x)       &
\end{cases}
 \end{eqnarray}
with $_1F_2(a;b,c;x)$ is the hypergeometric function, $\textrm{Re} (\Upsilon(x)) $ is the real part 
and $\textrm{Im} (\Upsilon(x)) $
is the imaginary part 
of the complex function $\Upsilon(x)$ given by
\begin{eqnarray}
\Upsilon(x) &=& 
(-\mathrm{i})^\alpha e^{-\mathrm{i}k \pi \frac{x}{2 q}}\Big( 
 \Gamma(-\alpha,-\mathrm{i} k \pi  \frac{q-x}{2 q}) -
 \Gamma(-\alpha,-\mathrm{i} k \pi  \frac{q+x}{2 q}) \Big)
\end{eqnarray}
where $\Gamma(a,z) $ denotes the incomplete $\Gamma$-function on the complex plane.

The kinetic part of the fractional Schr\"odinger equation may therefore be calculated analytically and may be written  with the testfunctions $^\infty\Psi^+(x)=\cos(k \pi/2 x)$ and $^\infty\Psi^-(x)=\sin(k \pi/2 x)$: 
\begin{eqnarray}
-{_\textrm{\tiny{RZ}}^\sqcup} D^\alpha \cos(k \pi/2 x) &=& E g_k(x)
\quad k=1,3,5,...\\
-{_\textrm{\tiny{RZ}}^\sqcup} D^\alpha \sin(k \pi/2 x) &=& E g_k(x)
\quad k=2,4,6,...
\end{eqnarray}
We define a pseudo-normalization condition 
\begin{eqnarray}
\label{cq_norm}
g_k(x=0)  &=& 1 \quad k=1,3,5,...\\
g_k(x=1/k)  &=& 1 \quad k=2,4,6,...
\end{eqnarray}
which allows to compare $g_k(x)$ with the trigonometric functions. 

In figures \ref{fcos} and \ref{fsin} the results are sketched. 
The bad news is, that neither $g_k(x) =  \cos(k \pi/2 x)$ nor $g_k(x) =  \sin(k \pi/2 x)$ holds. Consequently  the solutions of the free fractional Schr\"odinger equation based on  the Riesz derivative definition are no eigenfunctions of the same fractional Schr\"odinger equation with infinite potential well. This  answers the first question. 

The good news is, that the deviations become more and more negligible for $k \gg 1$. This answers the second question: For large k,   in the vicinity of $\alpha \approx 2$ and for $x \approx 0 $ the trigonometric functions seem to be  a good first guess.

As a consequence of the pseudo-normalization condition (\ref{cq_norm}), we can give an analytic expression for the approximate energy spectrum. With 
\begin{equation}
-{_\textrm{\tiny{RZ}}^\sqcup} D^\alpha \cos(k \pi/2 x) |_{x=0}= E^\sim_k(\alpha)
\quad k=1,3,5,...
\end{equation}
we obtain (since $I_2$  vanishes for $x=0$):
\begin{equation}
\label{etilde}
E^\sim_k(\alpha) = 
\frac{1}{2} q^{-\alpha}\Gamma(\alpha)
\frac{\sin(\frac{\pi}{2}\alpha)}{\pi}\Big(
\frac{k^2 \pi^2}{2-\alpha} \,
 _1F_2\big(1-\frac{\alpha}{2};\frac{3}{2}, 2-\frac{\alpha}{2}; -\frac{1}{16}k^2 \pi^2 \big) -
4 \cos(k \frac{\pi}{2})\Big)
\end{equation}
which according to the above mentioned criteria will turn out to be a good approximation of the exact energy eigenvalues for large k  in the vicinity of $\alpha \approx 2$. 

\setcounter{section}{2}
\setcounter{equation}{0}
\section{The solution - numerical part}
Since we have shown by explicit analytic evaluation of the fractional derivative, that the trigonometric functions are no eigenfunctions of the infinite potential well, but only good approximations, we will calculate the exact solutions numerically. 

For that purpose, we expand the exact solution $\Psi^\pm_k(x)$ in a Taylor series
\begin{eqnarray}
\Psi^+_k(x)  &=&\lim_{N \rightarrow \infty} \sum_{n=0}^N a_{2 n} x^{2 n} \\
\Psi^-_k(x)  &=& \lim_{N \rightarrow \infty} \sum_{n=0}^N a_{2 n+1} x^{2 n+1} 
\end{eqnarray}
and insert it into the fractional Schr\"odinger equation for the infinite well potential:
\begin{equation}
-{_\textrm{\tiny{RZ}}^\sqcup} D^\alpha \Psi^\pm_k(x) = E_k   \Psi^\pm_k(x)
\end{equation}

The integrals on the left may be evaluated analytically and lead to transcendental functions in x, which  then are  expanded up to order $N$ in a Taylor-series too. This leads to
\begin{eqnarray}
\sum_{n=0}^N b_{2 n} x^{2 n}  &=& E_k  \sum_{n=0}^N a_{2 n} x^{2 n}\\
 \sum_{n=0}^N b_{2 n+1} x^{2 n+1}&=& E_k   \sum_{n=0}^N a_{2 n+1} x^{2 n+1}
\end{eqnarray}
A term by term comparison results in a  system of non-linear equations of the type $\{E_k a_n  = b_n(a_n)\}$ on the set of variables $\{a_n,E_k\}$, which is  solved numerically.
For practical calculations we set  $N=20$, which yields an accuracy of the calculated energy levels of about $0.25\%$ for the ground state. 

Results are presented in figures \ref{fcosexact} and \ref{fsinexact}. For $\alpha=2$ we obtain the classical trigonometric functions. For decreasing $\alpha$ the eigenfunctions show a increasing tendency to shift towards the walls. This is exactly the behavior, which is not modeled by the trigonometric  test functions presented in the last section.  
In figure \ref{fenergiesexact} we compare the determined energy levels with the presented energy formulas. Especially  $E^\sim_k(\alpha)$ from (\ref{etilde}) is a useful approximation.

\section{Conclusion}
We have demonstrated, that the non-local character of the fractional operators used in the fractional Schr\"odinger equation  indeed needs special attention. Concepts like box normalization, WKB-approximation or piecewise solution may work well in a classical local approach, but cause errors when applied to  non-local problems. 

On the other hand, we have shown, that such local strategies may lead to useful approximations e.g. in low-level ($\alpha = 2- \epsilon $) fractional problems.  

This may be understood from 
\begin{equation}
\lim_{q \rightarrow \infty}{_\textrm{\tiny{RZ}}^\sqcup} D^\alpha  =  {_\textrm{\tiny{RZ}}^\infty} D^\alpha 
\end{equation}
as  a consequence of the specific weight $w(\xi)= \frac{1}{\xi^{\alpha+1}}$ in the integral definition of the Riesz fractional derivative  definition (\ref{q12driesz}) with the property
\begin{equation}
\lim_{\xi \rightarrow \infty}w(\xi) = 0
\end{equation}

The infinite potential well serves as a helpful tool to demonstrate the   consequences of different approaches.
\\
\\
\newpage
\section*{Acknowledgment}
We thank A. Friedrich for useful discussions.


\bigskip 
\bigskip 
\sl
\noindent
$^1$ gigaHedron\\
Berliner Ring 80 \\
D-63303 Dreieich, Germany  \\
e-mail: herrmann@gigahedron.com\\

\begin{thebibliography}{99}
%
\bibitem{Ab} 
Abramowitz, M. and  Stegun, I.~A.  (1965). 
\emph{Handbook of mathematical functions} 
Dover Publications, New York

\bibitem{bay12a} 
Bayin, S.~S. (2012).
\emph{On the consistency of the solutions of the space fractional Schr\"odinger equation}
arXiv:1203.4556v1 [math-ph],
J. Math. Phys. {\bf 53} 042105

\bibitem{cap67} Caputo, M. (1967)
 \emph{Linear model of dissipation whose Q is almost frequency independent Part II}
 Geophys. J. R. Astr. Soc  \textbf{13}, 529--539

\bibitem{don07} 
Dong, J. and Xu, M. (2007).
\emph{Some solutions to the space fractional Schr\"odinger equation using momentum representation method}
J. Math. Phys.  {\bf 48}, 072105

\bibitem{erd53}
 Erdelyi, A., Magnus, A., Oberhettinger, F. and Tricomy, F.~G. (1953)
\emph{Higher transcendental functions} Vol. II, Bateman Manuscript Project, 
California Institute of Technology, McGraw Hill, New York

\bibitem{feller} 
Feller, W. (1952).
\emph{On a generalization of Marcel Riesz' potentials and the semi-groups generated by them}
 Comm. Sem. Mathem. Universite de Lund,  72--81

\bibitem{guo06} 
Guo, X. and Xu, M.  (2006). 
\emph{Some physical applications of fractional Schr\"odinger equation}
J. Math. Phys.  \textbf{47},  082104

\bibitem{haw12} 
Hawkins, E. and Schwarz, J.~M. (2012).
\emph{Comment "On the consistency of solutions of the space fractional Schr\"odinger equation"}
arXiv:1210.1447 [math-ph]

\bibitem{her05}  
Herrmann, R.  (2005) 
\emph{Properties of a fractional derivative Schr\"odinger type wave equation and a new
interpretation of the charmonium spectrum} arXiv:math-ph/0510099

\bibitem{her11} Herrmann, R. (2011) \emph{Fractional calculus - an introduction for physicists},
World Scientific Publishing, Singapore



\bibitem{f2}
Liouville, J. (1832).
\emph{Sur le calcul des differentielles $\acute{\text{a}}$ indices quelconques}
 J. $\acute{\text{E}}$cole Polytechnique  {\bf 13}, 1--162

\bibitem{jen} 
Jeng, M., Xu, S.-L.-Y.,  Hawkins, E. and Schwarz J.~M. (2008).
\emph{On the non-locality of the fractional Schr\"odinger equation}
arXiv:0810.1543v1 [math-ph], 
J. Math. Phys. 51, 062102 (2010)

\bibitem{laskin} 
Laskin, N. (2002). 
\emph{Fractional Schr\"odinger equation}
Phys. Rev. E {\bf 66}, 056108--0561014

\bibitem{old76} Oldham, K.~B. and Spanier, J. (1976) \emph{The fractional calculus},
Dover Publications, Mineola, New York

\bibitem{pod99} Podlubny, I. (1999) \emph{Fractional differential equations},
Academic Press, New York

\bibitem{rie47}  
Riemann, B. (1847)
 \emph{ Versuch einer allgemeinen Auffassung der Integration
und Differentiation} in:  Weber, H. and Dedekind, R.  (Eds.) (1892)
\emph{Bernhard Riemann's gesammelte mathematische Werke und wissenschaftlicher Nachlass},  
Teubner, Leipzig, reprinted in
\emph{Collected works of Bernhard Riemann},  
Dover Publications (1953), 353--366

\bibitem{riesz} 
Riesz, M.  (1949). 
\emph{L'integrale de Riemann-Liouville et le probl$\acute{\text{e}}$me de Cauchy}
Acta Math.   \textbf{81},  1--222
\bibitem{weyl} 
Weyl, H. (1917).
\emph{Bemerkungen zum Begriff des Differentialquotienten gebrochener Ordnung}
Vierteljahresschr. Naturforsch. Ges. Z\"urich, {\bf{62}}, 296--302

\end{thebibliography}
\end{document}